\documentclass[10pt,journal,twocolumn,final]{IEEEtran}

\usepackage{eqnarray,amssymb,amsmath,amsthm,mathtools,amsfonts}
\usepackage{mathrsfs}
\usepackage[utf8]{inputenc}
\usepackage[T1]{fontenc}
\usepackage{graphicx}

\usepackage{subfigure}
\usepackage{color}
\usepackage{epstopdf}
\usepackage{cases}
\usepackage{color}
\usepackage{cite}
\usepackage{bbm}
\usepackage{algorithm}
\usepackage{algpseudocode}
\usepackage{soul}
\newtheorem{theorem}{Theorem}

\begin{document}
\title{Taming the Tail of Maximal Information Age in Wireless Industrial Networks}

\author{\IEEEauthorblockN{Chen-Feng Liu,~\IEEEmembership{Student Member,~IEEE}, and  Mehdi Bennis,~\IEEEmembership{Senior Member,~IEEE}}
\thanks{
This work was supported in part by the Academy of Finland project CARMA, in part by the Academy
of Finland project MISSION, in part by the Academy of Finland project SMARTER, in part by the INFOTECH project NOOR, in part by the Nokia Bell-Labs project ELLIS, in part by the Nokia Bell-Labs project UEBE, and in part by the Nokia Foundation.
}
\thanks{The authors are with the Centre for Wireless Communications, University of Oulu,  90014 Oulu, Finland (e-mail: chen-feng.liu@oulu.fi;  mehdi.bennis@oulu.fi).}
}

\maketitle

\begin{abstract}
In wireless industrial networks, the information of time-sensitive control systems needs to be transmitted in an ultra-reliable and low-latency manner. This letter studies the resource allocation problem in \emph{finite blocklength} transmission, in which the information freshness is measured as the \emph{age of information} (AoI) whose maximal AoI is characterized using \emph{extreme value theory} (EVT). The considered system design is to minimize the sensors' transmit power and transmission blocklength subject to constraints on the maximal AoI's tail behavior. The studied problem is solved using Lyapunov stochastic optimization, and a dynamic reliability and age-aware policy for resource allocation and status updates is proposed. Simulation results validate the effectiveness of using EVT to characterize the maximal AoI. It is shown that sensors need to send larger-size data with longer transmission blocklength at lower transmit power.
Moreover, the maximal AoI's tail decays faster at the expense of higher average information age.
\end{abstract}


\section{Introduction}
\label{sec:intro}

\IEEEPARstart{T}{he} timely delivery of control information in industrial Internet of things (IoT) with ultra reliability and low latency is of paramount importance \cite{CISS,5GACIA,TVT_finite,APCC}.  In industrial automation applications, e.g., process monitoring, motion control of packaging machines, and mobile crane control,  typical payload sizes range from 20 bytes to 250 bytes  \cite{5GACIA},
making a system design based on Shannon capacity (i.e., assuming infinite blocklength\footnote{Following reference \cite{finite_block}, we use the terminology ``blocklength'' to refer to the codeword length of the error-correcting code in channel coding.}) inadequate.  To address this issue, the \emph{finite-blocklength} transmission rate \cite{finite_block} has been considered in the wireless industrial setting  \cite{TVT_finite,APCC}. Nevertheless, an information decoding error occurs because the blocklength is not long enough to average out the effects of thermal noise and other distortions. 
At the central controller, the \emph{freshness} of the received status information is  important since any outdated information can potentially degrade the performance. However, to the best of our knowledge,  maintaining information freshness and optimizing the status updates under ultra-reliable low latency communication have received little attention within the industrial automation setting.
The main goal of this letter is to characterize the information freshness in ultra-reliable and low-latency industrial IoT, where multiple sensors wirelessly transmit their latest sampled status information to a central controller.
To measure the freshness of the status information, we leverage the notion of \emph{age of information} (AoI) \cite{AoIOrigin,V2V_AoI}, which is defined as \emph{the elapsed time since the status data was generated at the sensor until the current time instant}, while focusing on the maximal AoI over a given time interval.
Note that \emph{extreme value theory} (EVT) provides a powerful tool to investigate the asymptotic statistics of maximal metrics \cite{EVT:Han}.
Thus, we first characterize the maximal AoI using EVT and then impose time-averaged constraints on the maximal AoI's tail/decay behavior \cite{MehdiURLLC}. 
Moreover, in order to keep the central controller's received information as fresh as possible, sensors need to sample and update the status information more frequently at the cost of depleting their batteries.
The studied problem is cast as sensors' time-averaged transmit power minimization subject to the imposed constraints on the maximal AoI's tail behavior.
To deal with the time-averaged objective and constraints of the studied problem, we resort to Lyapunov stochastic optimization \cite{Neely/Stochastic} and propose a dynamic reliability and age-aware policy for sensors' transmit power, blocklength allocation, and status update.
Numerical results show several key tradeoffs: 1) To meet the transmission rate requirement, sensors require longer transmission blocklength and consume more energy, but use lower transmit power as data size increases; 2) More frequent status updates are required for denser networks; 3) A lower occurrence probability of extremely high information age is obtained at the expense of higher average age.
Moreover, we numerically verify the effectiveness of characterizing the tail behavior of the maximal AoI using EVT.

\section{System Model}
\label{sec:system}
Considering a wireless industrial network, we focus on the uplink transmission in which $K$ wireless sensors send their status information (e.g., for process monitoring in industrial automation) to a central controller.
During the entire communication timeline, the transmissions are indexed by $n\in\mathbb{Z} ^{+}$, and the time instant of transmission $n$ is denoted by $t_{n}$.
Given the initial time instant $t_{0}=0$, the time interval between two successive transmissions or status updates is $S_n=t_{n}-t_{n-1}$. Further, one sensor transmits its updated information in each transmission $n$. 
Note that a moderate number $K$ is considered in this work.\footnote{When the number of sensors grows significantly, multiple orthogonal resource blocks are required.}
As mentioned previously, the payload size in industrial automation is less than 250 bytes.
Then based on Shannon capacity, the blocklength with a 3\,dB signal-to-noise ratio is lower than 1200, making capacity-based design inappropriate \cite{finite_block}. Instead,
we consider the finite-blocklength transmission rate
$R_n^k(L_n,\epsilon)= \log_2(1+\gamma_n^k) -\frac{\sqrt{2\gamma_n^k(\gamma_n^k+2)}\mbox{erfc}^{-1}(2\epsilon)}{\sqrt{L_n}(1+\gamma_n^k)\ln 2},$
which incorporates the blocklength $L_n\ll\infty$ and a decoding error probability $\epsilon>0$, in the $n$-th transmission.
The notation $k$ represents the scheduled sensor in the $n$-th transmission. In addition, we have $\gamma_n^k=\frac{P_n^kh_n^k}{N_0 W}$ in which $P_n^k$ is the sensor's transmit power, and $h_n^k$ is the channel gain, including path loss and channel fading, between the sensor and controller.  
We also assume that the wireless channel experiences block fading and stays invariant in each transmission. $N_0$, $W$, and $\mbox{erfc}^{-1}(\cdot)$ are the power spectral density of the additive white Gaussian noise, bandwidth, and inverse complementary error function, respectively. 
For each transmission, the controller calculates the sensor's transmit power with $0\leq P_n^k\leq P^k_{\max}$, blocklength $L_n\in\mathbb{Z}^{+}$, and status-updating time interval with $S_n\geq S_{\min}$. Here, $P^k_{\max}$ and $S_{\min}$ are the sensor $k$'s power budget and the smallest interval value, respectively.
When the controller allocates the transmit power and blocklength, the rate requirement $R_n^k \geq  D^k/L_n$ has to be taken into account to ensure a sufficient rate for sending the status data with  size $D^k$. 
Let us denote the AoI (measured at the central controller) of sensor $k$'s monitored status data at any continuous time instant $t\geq 0$ as $\tau^k(t)$ and specify its details as follows. Firstly, once the sensor $k$ is granted access to the controller, it samples the status information and transmits it immediately. 
Since data size is small, we neglect the signal processing time and transmission time in the AoI calculation. 
Hence, provided that sensor $k$ correctly delivers the updated data in the $n$-th transmission, its AoI at time instant $t=t_n$ is reset to zero. Otherwise, the age increases by $S_n$. We formally express the AoI function as $\tau^k(t_{n})=(\tau^k(t_{n-1})+S_{n})(1-B_n^k\times \mathbbm{1}_{\{P_n^k>0\}})$
with the indicator function $\mathbbm{1}_{\{\cdot\}}$. 
Here, a Bernoulli random variable, $B_n^k\sim {\rm B}(1,1-\epsilon)$, is introduced to capture the success (by $B_n^k=1$) and failure (by $B_n^k=0$) of  information decoding. 
For any time instant between two successive transmissions, the AoI increases linearly as per  $\tau^k(t)=\tau^k(t_{n-1})+t-t_{n-1}$.

\section{Statistical Constraints on the Tail Behavior of the Maximal Age of Information}\label{Sec: Tail feature}

\begin{figure}[t]
\centering
	\includegraphics[width=\columnwidth]{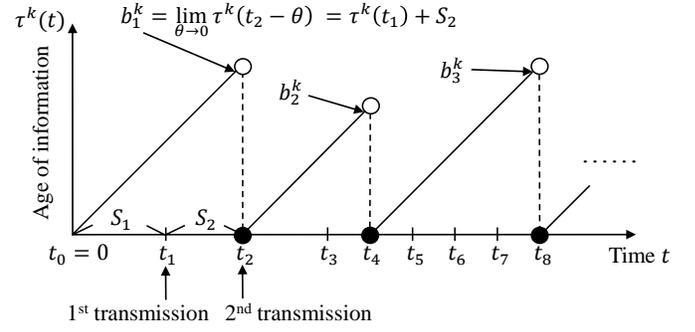}
	\caption{Illustrations of the transmission index $n$, transmission time instant $t_{n}$,   status-updating time interval $S_n$, AoI function $\tau^k(t)$, and peak AoI $b^k_m$. For instance, if sensor $k$ is scheduled at $n\in\{2,4,6,8\}$, then $B_n^k=1$ when $n=2, 4, 8$, and $B_n^k=0$ at $n=6$.}
		\label{Fig: AoI}
\end{figure}

As time elapses, the available data at the controller becomes outdated. This inaccurate status information can cause manufacturing failures or other adverse artifacts. Thus, we model the impact of information aging as a cost function $f(\cdot)$ and impose a constraint on the average cost (over all transmission time instants) of every sensor's AoI as
$\bar{f}^{\tilde{k}}\equiv \textstyle \lim\limits_{N\to\infty}\frac{1}{N}\sum_{n=1}^{N}\mathbb{E} [f(\tau^{\tilde{k}}(t_n))]\leq f^{\tilde{k}}_{{\rm th} },\forall\,\tilde{k}\in\mathcal{K}.$
Here, $f^{\tilde{k}}_{{\rm th} }$ is the threshold for the cost, and   $\tilde{k}$ refers to every sensor.
In addition to the AoI in each transmission, we are concerned with a worst-case metric $\max_{t\in[T_{i},T_{i+1})}\{\tau^{\tilde{k}}(t)\},\forall\,T_{i}\geq 0,i\in\mathbb{Z}^{+}$, namely, the maximal AoI over a time interval,  which can be explained as the ``oldest age'' of status information during the considered time period.
Given that the status update is successfully delivered to the controller in the $n$-th transmission,  we denote the peak AoI before the AoI is reset as
$b_m^k=\lim\limits_{\theta\to 0^{+}}\tau^k(t_n-\theta)=\tau^k(t_{n-1})+S_{n},$
where $m=\sum_{\tilde{n}=1}^{n}\mathbbm{1}_{\{\tau^k(t_{\tilde{n}})=0\}}$ represents the $m$-th successful delivery of sensor $k$'s status updates since $t=0$. From Fig.~\ref{Fig: AoI}, it can be straightforwardly seen that the maximal AoI over a time interval is equivalent to the maximum of all peak AoI within this interval, i.e., $\max_{t\in[T_{i},T_{i+1})}\{\tau^{\tilde{k}}(t)\}=\max_{m\in\{M_{i},\cdots,M_{i+1}-1\}}\{b_m^{\tilde{k}}\}$. Here, $M_i\in\mathbb{Z}^{+}$ is the corresponding order of successful information delivery.
As shown in Fig.~\ref{Fig: AoI}, the AoI is reset to zero after successful information delivery, in which the successes of decoding, i.e., $B_n^k$, are {\it i.i.d.}~in different transmissions. In addition, the wireless channel experiences block fading, so we can assume a stationary peak AoI process in which each peak AoI $b_m^{\tilde{k}},\forall\,m\in\mathbb{Z}^{+}$, has the same marginal distribution.
In order to further characterize the maximal AoI over a time interval, we next introduce some useful results of EVT in Theorem \ref{Thm: GEV}.
\begin{theorem}\label{Thm: GEV}
Let $X_1,X_2,\cdots$ be a stationary process with the same marginal distribution as a random variable $X$ and define  $Z_M\equiv\max\{X_1,\cdots,X_M\}$.  As $M\to\infty$, the complementary cumulative distribution function (CCDF) of  $Z_M$ converges to a generalized extreme value (GEV) distribution whose statistics are characterized by a location parameter $\mu\in\mathbb{R}$, a scale parameter $\sigma >0$, and a shape parameter $\xi \in\mathbb{R}$ \cite{EVT:Han}.
\end{theorem}
By applying Theorem \ref{Thm: GEV} to the stationary peak AoI process $\{b_m^{\tilde{k}}\}$, the maximal AoI, i.e., $\max_{m\in\{M_{i},\cdots,M_{i+1}-1\}}\{b_m^{\tilde{k}}\}$, can be characterized by a GEV distribution when $(M_{i+1}-M_{i})\to\infty$.
Furthermore, governed by the shape parameter $\xi$ for the tail/decay behavior of the CCDF, the GEV distributions are categorized into three types according to the shape parameter $\xi$ \cite{EVT:Han}: 
\emph{(i)} When $\xi<0$, the GEV distribution  is \emph{short-tailed}, having a finite endpoint at $z_{\rm end}=\bar{F}^{-1}_Z(0)= \mu-\sigma/\xi<\infty$ in the CCDF. \emph{(ii)} The CCDF  has a thinner tail than an exponential function when $\xi=0$. In this case, the GEV distribution is \emph{light-tailed}.  \emph{(iii)} When $\xi>0$, the GEV distribution whose tail is more weighted than an exponential function is \emph{heavy-tailed}. In the latter two types,  the endpoints of the CCDFs approach infinity, i.e., $z_{\rm end}=\bar{F}^{-1}_Z(0)\to\infty$.
Now let us impose a constraint on the tail behavior of the approximated GEV distribution of the maximal AoI, e.g., $\xi^{\tilde{k}}_{\rm th}<0,\forall\,\tilde{k}\in\mathcal{K}$.  For tractable analysis, we transform $\xi^{\tilde{k}}_{\rm th}<0$ to another expression by resorting to  the Pickands–Balkema–de Haan theorem  \cite{EVT:Han}, in which the shape parameter in Theorem \ref{Thm: GEV} is $\xi=\frac{\mathbb{E}[(X-q)^2|X>q]-2\mathbb{E}[X-q|X>q]^2}{2{\rm Var}(X-q|X>q)}$ given a threshold $q$ with $F_{X}(q)\approx 1$.
Incorporating the expression of $\xi$ and the constraint $\xi^{\tilde{k}}_{\rm th}<0$,
we consider the following two statistical constraints for the tail behavior of sensor $\tilde{k}$'s maximal AoI, i.e., $\bar{Y}^{\tilde{k}}-\delta\geq\eta^{\tilde{k}}$ and $2(\eta^{\tilde{k}})^2\geq\bar{\Upsilon}^{\tilde{k}}+\delta$, with a predetermined value $\eta^{\tilde{k}}$ and an infinitesimal positive value $\delta$.
Here,  $\bar{Y}^{\tilde{k}}\equiv\lim\limits_{M\to\infty}\frac{1}{M}\sum_{m=1}^{M}\mathbb{E}\big[Y^{\tilde{k}}_m|b_m^{\tilde{k}}> q^{\tilde{k}} \big]$, $\bar{\Upsilon}^{\tilde{k}}\equiv\lim\limits_{M\to\infty}\frac{1}{M}\sum_{m=1}^{M}\mathbb{E}\big[(Y^{\tilde{k}}_m)^2 |b_m^{\tilde{k}}> q^{\tilde{k}}\big]$, $Y^{\tilde{k}}_m=b_m^{\tilde{k}}- q^{\tilde{k}}$,  and $q^{\tilde{k}}\approx F^{-1}_{b_m^{\tilde{k}}}(1)$.
%
%
%
Our studied optimization problem is formulated as
\begin{subequations}\label{Eq: main problem}
\begin{IEEEeqnarray}{cll}
\hspace{-1.5em}\underset{P^k_n,L_n,S_n}{\mbox{minimize}}&~~\lim\limits_{N\to\infty}\frac{1}{N}\sum\limits_{n=1}^{N} \frac{P_n^kL_n}{S_nW}&\label{Eq: main problem-1}
\\\hspace{-1.5em}\mbox{subject to}&~~R_n^k\geq   D^k/L_n,&~~\forall\,n\in\mathbb{Z}^+,\label{Eq: main problem-2}
\\&~~\bar{f}^{\tilde{k}}\leq f^{\tilde{k}}_{{\rm th}},&~~\forall\,\tilde{k}\in\mathcal{K}, \label{Eq: main problem-5}
\\&~~\bar{Y}^{\tilde{k}}-\delta\geq \eta^{\tilde{k}},~
2(\eta^{\tilde{k}})^2\geq\bar{\Upsilon}^{\tilde{k}}+\delta,&~~\forall\,\tilde{k}\in\mathcal{K},\label{Eq: main problem-6}
\\&~~0\leq P_n^k\leq P^k_{\max},&~~\forall\,n\in\mathbb{Z}^+,\label{Eq: main problem-3}
\\&~~L_n\in\mathbb{Z}^{+},&~~\forall\,n\in\mathbb{Z}^+,\label{Eq: main problem-7}
\\&~~ S_n\geq S_{\min},&~~\forall\,n\in\mathbb{Z}^+.\label{Eq: main problem-4}
\end{IEEEeqnarray}
\end{subequations}
The objective function $\frac{P_n^kL_n}{S_nW}$ is the sensor $k$'s normalized power consumption with the transmission time length $L_n/W$. In the next section, Lyapunov stochastic optimization is used to solve problem \eqref{Eq: main problem}.

\section{Reliability and Age-Aware Resource Allocation and Status Update}

In the derivation procedures of Lyapunov optimization, we first introduce
three virtual queues
$Q_{\rm (f)}^{\tilde{k}}(n+1)=\max\big\{Q_{\rm (f)}^{\tilde{k}}(n)+f(\tau^{\tilde{k}}(t_n))- f^{\tilde{k}}_{{\rm th} },0\big\}$,
$Q_{\rm (m)}^{\tilde{k}}(m+1)=\max\big\{ Q_{\rm (m)}^{\tilde{k}}(m)-(Y^{\tilde{k}}_m-\eta^{\tilde{k}}-\delta)\times \mathbbm{1}_{\{b_m^{\tilde{k}}> q^{\tilde{k}}\}},0\big\}$,
and
$Q_{\rm (v)}^{\tilde{k}}(m+1)=\max\big\{Q_{\rm (v)}^{\tilde{k}}(m)+\big[(Y^{\tilde{k}}_m)^2-2(\eta^{\tilde{k}})^2+\delta\big]\times \mathbbm{1}_{\{b_m^{\tilde{k}}> q^{\tilde{k}}\}},0\big\}$
for each long-term average constraint in \eqref{Eq: main problem-5} and \eqref{Eq: main problem-6}.\footnote{The transformed statistical constraints and corresponding virtual queues for the cases $\xi^{\tilde{k}}_{\rm th}=0$ and $\xi^{\tilde{k}}_{\rm th}>0$ are shown in the Appendix.}
Due to space limitations, we skip the intermediate derivations and directly show the results after applying Lyapunov optimization. That is, at each transmission $n\in\mathbb{Z}^{+}$, the controller solves
\begin{IEEEeqnarray*}{rcl}
\mbox{{\bf MP:}}~~&\underset{P^k_n,L_n,S_n}{\mbox{minimize}}&~~\phi_nS_{n}+\psi_ne^{S_n}+\frac{VP_n^{k}L_n}{S_nW}
\\&\mbox{subject to}&~~\mbox{\eqref{Eq: main problem-2}, \eqref{Eq: main problem-3}, \eqref{Eq: main problem-7}, and \eqref{Eq: main problem-4}},
\end{IEEEeqnarray*}
considering the exponential cost function $f(\cdot)=e^{(\cdot)}$ and assuming $B_n^k=1$ for the scheduled sensor $k$. Here,\footnote{$\phi_n$ for the cases $\xi^{\tilde{k}}_{\rm th}=0$ and $\xi^{\tilde{k}}_{\rm th}>0$ is shown in the Appendix.} $\phi_n=2Q_{\rm (v)}^k(m)\tau^k(t_{n-1})+2[\tau^k(t_{n-1})]^3-Q_{\rm (m)}^k(m)\tau^k(t_{n-1})-Q_{\rm (m)}^k(m)$ and  $\psi_n=\sum_{\tilde{k}\in\mathcal{K}\setminus k}Q_{\rm (f)}^{\tilde{k}}(n) e^{\tau^{\tilde{k}}(t_{n-1})}$.
To solve {\bf MP}, we first fix the variable $S_n$ and obtain a sub-problem {\bf SP1} in which the optimal transmit power and blocklength, denoted by $P^{k*}_n(S_n)$ and $L^{*}_n(S_n)$, are functions of $S_n$ in general. 
Subsequently, pluging $P^{k*}_n(S_n)$ and $L^{*}_n(S_n)$ into {\bf MP}, we have the second sub-problem {\bf SP2} which gives the optimal status-updating time interval $S^{*}_n$. The details of {\bf SP1} and {\bf SP2} are shown as follows:
\begin{IEEEeqnarray}{rl}
\mbox{{\bf SP1:}}~~&\underset{P^k_n,L_n}{\mbox{minimize}}~~P^k_nL_n
~~\mbox{subject to}~\mbox{\eqref{Eq: main problem-2}, \eqref{Eq: main problem-3}, and \eqref{Eq: main problem-7}},\notag
\end{IEEEeqnarray}
%
%
where $S_n$ is removed without affecting the solution. In other words,  $P_n^{k*}$ and $L_n^*$ are the constant functions of $S_n$.
\begin{IEEEeqnarray*}{rcl}
\mbox{{\bf SP2:}}~~&\underset{S_n\geq S_{\min}}{\mbox{minimize}}&~~\phi_nS_{n}+\psi_ne^{S_n}+\frac{VP_n^{k*}L_n^*}{S_nW}
\end{IEEEeqnarray*}
whose optimal solution, obtained by differentiation, is
$S_n^*=\max\{S_{\min},\tilde{S}_n\}$ with $\tilde{S}_{n}$ satisfying $\psi_ne^{\tilde{S}_{n}}+\phi_n=\frac{V P_n^{k*}L_n^*}{(\tilde{S}_n)^2W}$.
In the following derivations, let us solve the mixed-integer non-convex optimization problem {\bf SP1}, where
the subscript $n$ is neglected for notational simplicity. We first relax $L\in\mathbb{Z}^+$ as $L\geq 1$ and  introduce an auxiliary variable vector $\boldsymbol{\alpha}\equiv(\varsigma,\eta,a,b,g,\rho)\in\mathbb{R}^6$ which satisfies
\begin{subequations}\label{Eq: variable transformation}
\begin{IEEEeqnarray}{rlrl}
   e^{ \varsigma}\,&\leq  L,&
  \quad e^{\eta}\,&\leq N_0 W+ P^{k}h^{k},\label{Eq: variable transformation-1}
 \\ P^kh^k\, &\leq  e^{a},&
 \quad 2N_0 W+ P^{k}h^{k}\,&\leq e^{b},\label{Eq: variable transformation-2}
\\ P^k\,& \leq  e^{g}  ,&\quad L \,&\leq  e^{\rho}.\label{Eq: variable transformation-3}
\end{IEEEeqnarray}
\end{subequations}
Applying \eqref{Eq: variable transformation-1} and \eqref{Eq: variable transformation-2} to $ R^k$,  we obtain
$ R^k\geq \tilde{R}^k \equiv\log_2 (1+ \frac{P^{k}h^{k}}{N_0 W})- \frac{\sqrt{2}{\rm erfc}^{-1}(2\epsilon) e^{(a/2+b/2-\eta -\varsigma/2)}}{\ln 2}.$
If we can ensure $\tilde{R}^k \geq D^k/L$, the rate requirement \eqref{Eq: main problem-2} will be satisfied.
In addition, \eqref{Eq: variable transformation-3} provides an upper bound for the objective of {\bf SP1}, i.e., $P^kL\leq e^{(g+\rho)}$.
Thus, incorporating the above auxiliary variables into {\bf SP1}, we rewrite the optimization problem as
\begin{IEEEeqnarray*}{rcl}
\mbox{{\bf RP:}}~~&\underset{P^k,L,\boldsymbol{\alpha}}{\mbox{minimize}}&~~\textstyle g+\rho
\\&\mbox{subject to}&~~\tilde{R}^k\geq  D^k/L,~L\geq 1,~\mbox{\eqref{Eq: main problem-3}, and \eqref{Eq: variable transformation-1}--\eqref{Eq: variable transformation-3}}.\notag
\end{IEEEeqnarray*}
Note that solving {\bf RP} provides a sub-optimal solution for {\bf SP1}.
Due to the concave nature of \eqref{Eq: variable transformation-2} and \eqref{Eq: variable transformation-3} while the objective and  the remaining  constraints are affine and convex functions,
{\bf RP} belongs to the difference of convex programming problems. By iteratively convexifying the concave functions, the convex-concave procedure (CCP) provides a tractable approach  to solve {\bf RP}. Specifically, in the $j$-th iteration, we convexify \eqref{Eq: variable transformation-2} and \eqref{Eq: variable transformation-3} by the first-order Taylor series expansion
with respect to a reference point $\hat{\boldsymbol{\alpha}}_j$  and obtain the convexified optimization problem as
\begin{subequations}
\begin{IEEEeqnarray}{rcl}\label{CCP}
\hspace{-2em}\mbox{{\bf CP-$j$:}}~~&\underset{P^k,L,\boldsymbol{\alpha}}{\mbox{minimize}}&~~ g+\rho\label{CCP-1}
\\&\mbox{subject to}
&~~P^k h^k \leq  e^{\hat{a}_j}(a-\hat{a}_j) + e^{\hat{a}_j},\label{CCP-2}
\\&&~~2N_0W+P^k h^k\leq  e^{\hat{b}_j}(b-\hat{b}_j) +e^{\hat{b}_j},\label{CCP-3}
\\&&~~P^k\leq  e^{\hat{g}_j}(g-\hat{g}_j) +e^{\hat{g}_j}  ,\label{CCP-4}
\\&&~~L\leq e^{\hat{\rho}_j}(\rho-\hat{\rho}_j)  +e^{\hat{\rho}_j},\label{CCP-5}
\\&&~~\tilde{R}^k \geq D^k/L,~L\geq 1,~\mbox{\eqref{Eq: main problem-3}, and \eqref{Eq: variable transformation-1}.}\notag
\end{IEEEeqnarray}
\end{subequations}
Subsequently, the optimal solution to {\bf CP-$j$} is set as the reference point $\hat{\boldsymbol{\alpha}}_{j+1}$ of the next iteration.
Since providing the closed-form solution expression for problem {\bf CP-$j$} is not feasible,
we resort to CVX to numerically solve it.
Finally, using $(P^{\star}_{\infty},L^{\star}_{\infty})$ of the converged solution, we set sensor $k$'s transmit power and blocklength in the $n$-th transmission as $P^{k*}_n=P^{\star}_{\infty}$ and $L^*_n=\lceil L^{\star}_{\infty} \rceil$. After receiving the status information, the controller updates the sensors' AoI and the virtual queue values $Q_{\rm (f)}^{\tilde{k}}$,  $Q_{\rm (m)}^{\tilde{k}}$, and $Q_{\rm (v)}^{\tilde{k}}$.
The steps of the CCP algorithm are shown in Algorithm \ref{Alg: CCP} while
the procedures of the proposed resource allocation and status update mechanism are outlined in Algorithm \ref{Alg: Main mechanism}.

\begin{algorithm}[t]
 \caption{CCP for Solving {\bf RP}}
  \begin{algorithmic}[1]
  \State Initialize a feasible point $\hat{\boldsymbol{\alpha}}_1$ of {\bf RP} and  $j=1$.
    \Repeat
      \State Convexify  \eqref{Eq: variable transformation-2} and \eqref{Eq: variable transformation-3} by \eqref{CCP-2}--\eqref{CCP-5}.
      \State Solve problem {\bf CP-$j$} and denote the optimal solution as $(P_{j}^{\star}, L_{j}^{\star},\boldsymbol{\alpha}^{\star}_{j})$.
      \State Update $\hat{\boldsymbol{\alpha}}_{j+1}=\boldsymbol{\alpha}^{\star}_{j}$ and $j\leftarrow j+1.$
    \Until{Stopping criteria are satisfied.}
  \end{algorithmic}\label{Alg: CCP}
\end{algorithm}
 \begin{algorithm}[t]
  \caption{Reliability and Age-Aware Resource Allocation and Status Update Mechanism}
  \begin{algorithmic}[1]
  \State Initialize $m=0$ for each $k\in\mathcal{K}$ and $n=1$, predetermine the system lifetime as $N$, and set the initial queue values 
$Q_{\rm (f)}^{\tilde{k}}$, $ Q_{\rm (m)}^{\tilde{k}}$, and $Q_{\rm (v)}^{\tilde{k}},\forall\,\tilde{k}\in\mathcal{K}$, as zero.
      \Repeat
    \State Find transmit power $P^{k*}_n=P^{\star}_{\infty}$ and blocklength $L^*_n=\lceil L^{\star} _{\infty}\rceil$ by following Algorithm \ref{Alg: CCP} and the status-updating time interval $S_n^*$ by solving {\bf SP2}.
       \If{$B_n^k=1$}
          \State Update the queue lengths  $ Q_{\rm (m)}^{k}(m+1)$ and $Q_{\rm (v)}^{k}(m+1)$.
          \State    Update $m\gets m+1$ for the scheduled sensor $k$.
    \EndIf
     \State  Update  the queue lengths $Q_{\rm (f)}^{\tilde{k}}(n+1),\forall\,\tilde{k}\in\mathcal{K}$.
\State    Update $n\gets n+1$.
\Until{$n>N$}
  \end{algorithmic}\label{Alg: Main mechanism}
\end{algorithm}

\section{Numerical Results}
\label{sec:num}

We simulate the communication environment in a factory, considering the path loss model $33 \log x + 20 \log 2.625 + 32$ (dB) at the 2.625\,GHz carrier frequency \cite{rpt:itu_indoor}. The distance between the sensor and central controller is $x=15$\,m. In addition, the wireless channel experiences Rayleigh fading with unit variance. The remaining simulation parameters are $N_0=-174$\,dBm/Hz,  $W=1$\,MHz, $K\in\{2,3,\cdots,10\}$,  $D^k\in[20,250]$ bytes, $P_{\max}^{k}= 0$\,dBm, $V=1$, $S_{\min}=0$,  $\epsilon\in\{10^{-9},10^{-5}\}$, $f^{\tilde{k}}_{\rm th}=1.03$, $\eta^{\tilde{k}}=0.02$, $\delta=10^{-9}$, and $q^{\tilde{k}}= F^{-1}_{b_m^{\tilde{k}}}(0.99)$ \cite{CISS,5GACIA,TVT_finite,APCC,Liu_CL18}.

\begin{figure}[t]
\centering
\includegraphics[width=\columnwidth]{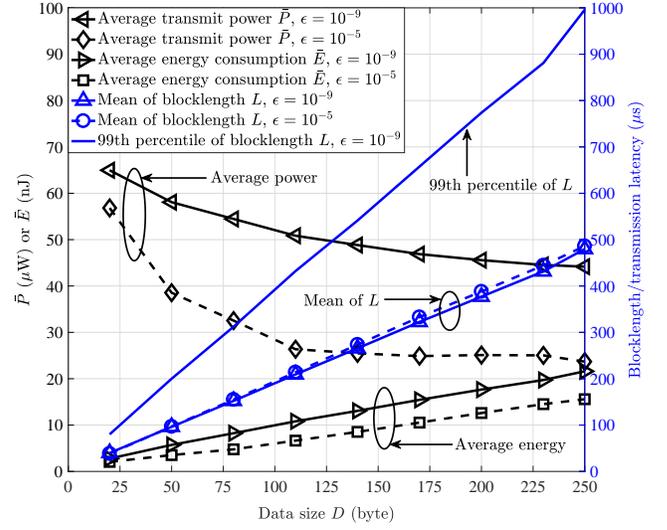}
\caption{1) Average transmit power or energy consumption in each transmission (left y-axis), and 2) mean and 99th percentile of the blocklength/transmission latency (right y-axis), versus data size.}
\label{Fig: P_L_D_tradeoff}
\end{figure}
\begin{figure}
\centering
\includegraphics[width=\columnwidth]{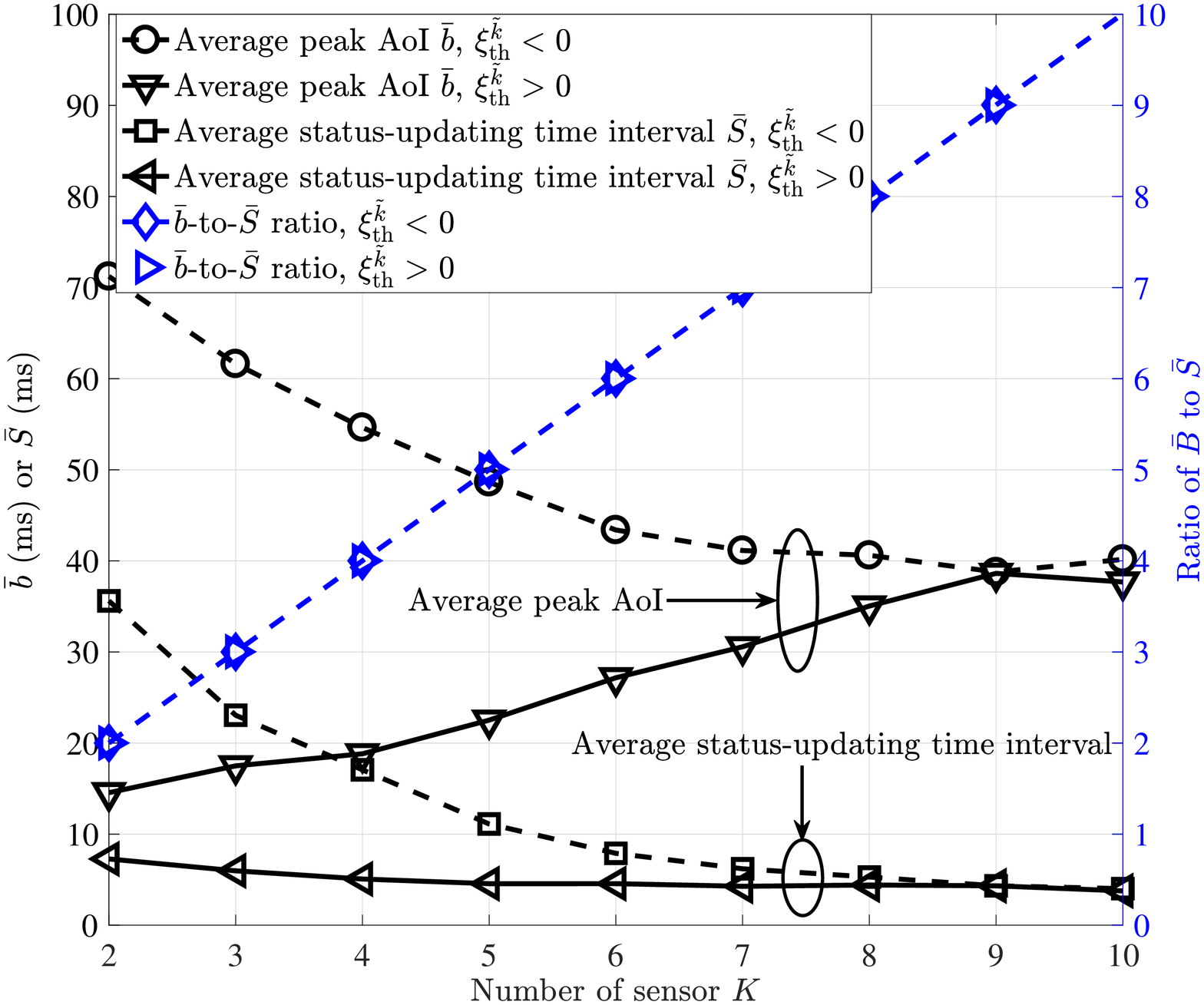}
\caption{Average peak AoI, average status-updating time interval, and their ratio versus number of sensors.}
\label{Fig: Sampling_AoI}
\end{figure}

Fig.~\ref{Fig: P_L_D_tradeoff} shows how the data size impacts the average transmit power, energy consumption in each transmission, transmission blocklength, and the 99th percentile blocklength. Note that the results in Fig.~\ref{Fig: P_L_D_tradeoff} are not affected by the number of sensors $K$. Intuitively, larger data sizes consume more energy for transmission, irrespective of the decoding error probability $\epsilon$. While we are concerned with energy minimization in problem {\bf SP1}, increasing energy transmission (for sending a larger data) by increasing the blocklength is more efficient given any specific value of $\epsilon$. In this regard, we can simply rewrite the rate requirement \eqref{Eq: main problem-2} as $D^k\leq L_nR_n^{k}\propto L_n\log_2(1+P_n^k)-\sqrt{L_n}\mbox{erfc}^{-1}(2\epsilon)$ in which the data size varies polynomially with the blocklength and logarithmically with transmit power. 
Therefore, the blocklength increases with the data size (and energy consumption), but the transmit power behaves oppositely. For different decoding error probabilities $\epsilon$, the blocklength and transmission rate are identical given a fixed data size. However, since $\mbox{erfc}^{-1}(2\epsilon)$ is smaller at $\epsilon=10^{-5}$, the sensor consumes less power and energy to achieve the same transmission rate. 

Considering $D^k=20$ bytes and $\epsilon=10^{-9}$ in Fig.~\ref{Fig: Sampling_AoI}, we show the average peak AoI and average status-updating time intervals for various network densities. In order to reduce the impact of information aging when the number of sensors grows, the controller decreases the interval between two successive transmissions, i.e., updates. 
\begin{figure}[t]
    \centering
   \subfigure[$\xi^{\tilde{k}}_{\rm th}<0$]{\includegraphics[width=\columnwidth]{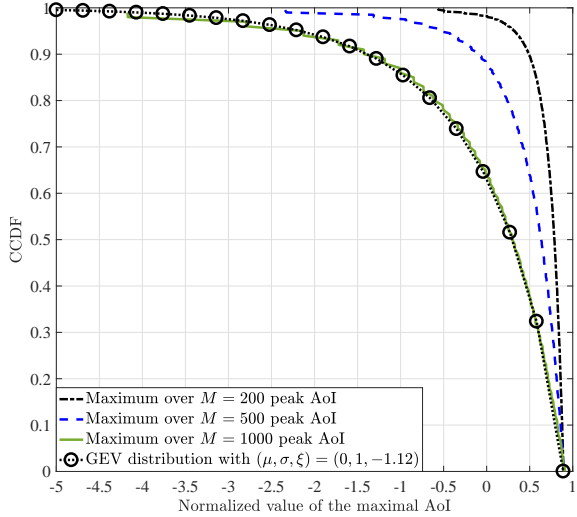}}
\subfigure[$\xi^{\tilde{k}}_{\rm th}>0$]{\includegraphics[width=\columnwidth]{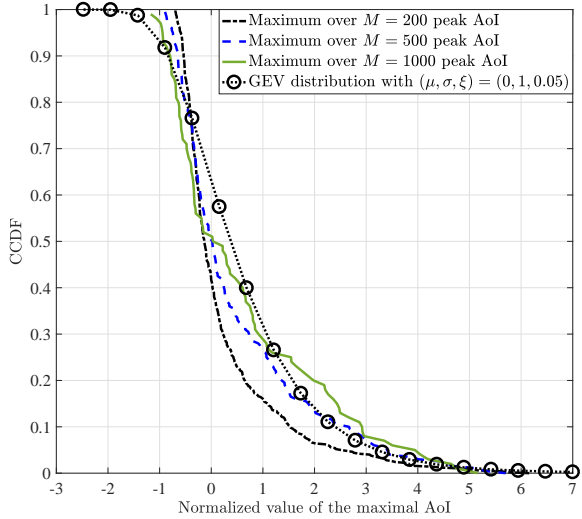}}
\caption{CCDFs of the normalized block maxima of peak AoI and the theoretically approximated GEV distribution.}
\label{Fig: AoI_Tail}
\end{figure}
Additionally, the average peak AoI is plainly the product of the number of sensors and the average status-updating time interval, i.e., $\bar{b}=K\bar{S}$. Additionally, in Figs.~\ref{Fig: P_L_D_tradeoff} and \ref{Fig: Sampling_AoI}, we find that the largest 99th quantile of transmission time and the lowest status-updating time interval are 1\,ms and around 4\,ms, respectively, in the simulated networks. Thus, when one transmission happens over a long time, the network still can have a sufficient time margin between successive updates.
Let us further fix $K=2$. Referring to the von Mises conditions and the runs estimator for the extremal index \cite{EVT:Han}, we normalize the maximum of $M$ (i.e., the block size) successive peak AoI values and show the CCDFs of normalized block maxima and the theoretically approximated GEV distribution in Fig.~\ref{Fig: AoI_Tail}. We can see that the numerical results converge to the approximated GEV distribution when the block size $M$ increases. In other words, these two figures verify the effectiveness of leveraging EVT to characterize the tail behavior of the maximal AoI in the considered wireless industrial network.
Furthermore, in contrast with the case $\xi^{\tilde{k}}_{\rm th}>0$, we have, in general, larger peak AoI (as shown in Fig.~\ref{Fig: Sampling_AoI}) but  smaller occurrence probability of extremely large peak AoI values in the case $\xi^{\tilde{k}}_{\rm th}<0$. This phenomenon means that the CCDFs of the normalized block maxima in Fig.~\ref{Fig: AoI_Tail}(a) are short-tailed, i.e., having a finite endpoint, whereas the curves in Fig.~\ref{Fig: AoI_Tail}(b) have heavy tails in which the endpoints approach infinity.

\section{Conclusions}
\label{sec:con}

In this letter, we have jointly taken into account reliability and information freshness to optimize sensors' status updates in wireless industrial networks. While allocating transmit power and transmission blocklength for sensors, we have taken into account the decoding error due to finite blocklength transmission. 
In addition, we have modeled the freshness of available information by its age and further characterized  the  statistics of the maximal AoI by using EVT.
Taking into account the constraint on the tail behavior of the maximal AoI, we have formulated the studied optimization problem  as sensors' transmit power minimization. 
Subsequently, we have proposed a dynamic reliability and age-aware policy for resource allocation and status updates.
Simulation results have shown that longer transmission blocklength with lower transmit power are required for delivering larger-size data, and have validated the effectiveness of using EVT to characterize the tail behavior of the maximal AoI. Finally, we have shown that the smaller occurrence probability of extremely high peak AoI is obtained at the cost of higher average peak AoI. 
In our future work, we will leverage statistical learning techniques to enhance reliability.

\appendix

For $\xi^{\tilde{k}}_{\rm th}=0$, we impose the constraints 
$\bar{Y}^{\tilde{k}}=\eta^{\tilde{k}}$ and $\bar{\Upsilon}^{\tilde{k}}=2(\eta^{\tilde{k}})^2$ in \eqref{Eq: main problem-6}.
The corresponding virtual queues are
$Q_{\rm (m)}^{\tilde{k}}(m+1)= Q_{\rm (m)}^{\tilde{k}}(m)+(Y^{\tilde{k}}_m-\eta^{\tilde{k}})\times \mathbbm{1}_{\{b_m^{\tilde{k}}> q^{\tilde{k}}\}}$ and $Q_{\rm (v)}^{\tilde{k}}(m+1)=Q_{\rm (v)}^{\tilde{k}}(m)+[(Y^{\tilde{k}}_m)^2-2(\eta^{\tilde{k}})^2]\times \mathbbm{1}_{\{b_m^{\tilde{k}}> q^{\tilde{k}}\}}$, respectively.
We have $\phi_n=2Q_{\rm (v)}^k(m)\tau^k(t_{n-1})+2[\tau^k(t_{n-1})]^3+2\tau^k(t_{n-1})+Q_{\rm (m)}^k(m)\tau^k(t_{n-1})+Q_{\rm (m)}^k(m)$ in problem {\bf MP}.

For $\xi^{\tilde{k}}_{\rm th}>0$, the imposed constraints in \eqref{Eq: main problem-6} are
$\bar{Y}^{\tilde{k}}+\delta\leq\eta^{\tilde{k}}$ and $2(\eta^{\tilde{k}})^2\leq\bar{\Upsilon}^{\tilde{k}}-\delta$ with the corresponding virtual queues
$Q_{\rm (m)}^{\tilde{k}}(m+1)= \max\{Q_{\rm (m)}^{\tilde{k}}(m)+(Y^{\tilde{k}}_m-\eta^{\tilde{k}}+\delta)\times \mathbbm{1}_{\{b_m^{\tilde{k}}> q^{\tilde{k}}\}},0\}$
and
$Q_{\rm (v)}^{\tilde{k}}(m+1)=\max\{Q_{\rm (v)}^{\tilde{k}}(m)-[(Y^{\tilde{k}}_m)^2-2(\eta^{\tilde{k}})^2-\delta]\times \mathbbm{1}_{\{b_m^{\tilde{k}}> q^{\tilde{k}}\}},0
\}$, respectively. In problem {\bf MP}, we have
 $\phi_n=-2Q_{\rm (v)}^k(m)\tau^k(t_{n-1})+2[\tau^k(t_{n-1})]^3+2\tau^k(t_{n-1})+Q_{\rm (m)}^k(m)\tau^k(t_{n-1})+Q_{\rm (m)}^k(m)$.

\bibliographystyle{IEEEtran}
\bibliography{Ref_IIoT_Lett}		

\end{document}